\documentclass[superscriptaddress,twocolumn]{revtex4}
\usepackage{amssymb}
\usepackage[tbtags]{amsmath}
\usepackage{graphicx}
\date{\today}

\begin{document}

\title{Coupling Josephson qubits via a current-biased information bus}
\author{L.F. Wei}
\affiliation{Frontier Research System, The Institute of Physical
and Chemical Research (RIKEN), Wako-shi, Saitama, 351-0198, Japan}
\affiliation{Institute of Quantum Optics and Quantum Information,
Department of Physics, Shanghai Jiaotong University, Shanghai
200030, P.R. China}
\author{ Yu-xi Liu}
\affiliation{Frontier Research System, The Institute of Physical
and Chemical Research (RIKEN), Wako-shi, Saitama, 351-0198, Japan}
\author{Franco Nori}
\affiliation{Frontier Research System, The Institute of Physical
and Chemical Research (RIKEN), Wako-shi, Saitama, 351-0198, Japan}
\affiliation{Center for Theoretical Physics, Physics Department,
CSCS, The University of Michigan, Ann Arbor, Michigan 48109-1120}

\begin{abstract}
Josephson qubits without direct interaction can be effectively
coupled by sequentially connecting them to an information bus: a
current-biased large Josephson junction treated as an oscillator
with adjustable frequency. The coupling between any qubit and the
bus can be controlled by modulating the magnetic flux applied to
that qubit. This tunable and selective coupling provides two-qubit
entangled states for implementing elementary quantum logic
operations, and for experimentally testing Bell's inequality.

phenomena, and Josephson effects.


paradox, Bell's inequalities, GHZ states, etc.).
\end{abstract}

\maketitle

Superconducting circuits with Josephson junctions offer one of the
most promising candidates for realizing quantum
computation~\cite{Makhlin01,Tsai99,Mooij99,Vion02,martinis02,Yu02,
Makhlin99,Tsai03,Averin03,You02,Blais03,Strauch03,Falci03,Zhu03,
Clarke88,Storcz03}.
These superconducting qubits can be either charge-~\cite{Tsai99},
flux-~\cite{Mooij99}, mixed-~\cite{Vion02}, current-biased
Josephson-junction (CBJJ) qubits~\cite{martinis02,Yu02}, and
others. Much attention is now devoted to realizing controlled
couplings between superconducting qubits and implementing quantum
logic operations~(see, e.g.,
\cite{Makhlin99,Tsai03,Averin03,You02,Blais03}). Two qubits, $i$
and $j$, can be connected by a common inductor or capacitor, with
Ising-type couplings
$\sigma^{(i)}_\alpha\otimes\sigma^{(j)}_\alpha\,(\alpha=x,y,{\rm
or}\,z)$. However, in general, (1) the capacitive
coupling~\cite{Tsai03,Strauch03} between qubits is not tunable
(and thus adjusting the physical parameters for realizing
two-qubit operation is not easy), and (2) a large inductance is
required in~\cite{Makhlin99} to achieve a reasonably high
interaction strength and speed for two-qubit
operations~\cite{You02}.
Alternatively, other schemes (see,
e.g.,~\cite{Falci03,Blais03,Zhu03}) use sequential interactions of
individual qubits with an information bus. These provide some
advantages: allow faster two-qubit operations, may possess longer
decoherence times, and are scalable. These schemes are similar to
the techniques used for trapped ions~\cite{CZ95}, where the ions
are entangled by exciting and de-exciting quanta (data bus) of
their shared collective vibrational modes.

Compared to the externally-connected $LC$-resonator used in
Ref.~\cite{Falci03} and the cavity QED mode proposed in
Ref.~\cite{Zhu03}, a large (e.g., $10\,\mu$m)
CBJJ~\cite{Clarke88,martinis02} is more suitable as an information
bus, because its eigenfrequency can be controlled by adjusting the
applied bias current.
In fact, such data bus to couple distant qubits has been proposed
in \cite{Blais03}. However, there all non-resonant interactions
between the qubits and the bus were ignored. This is problematic
because these near-resonance interactions must be considered,
otherwise, the desired coupling/decoupling between the chosen
qubit and the bus cannot be implemented because a perfect
resonance condition is not always achievable.
Also, modulating the bias current to selectively couple different
qubits changes the physical characteristics (e.g., eigenfrequency)
of the bus, and thus may yield additional errors during the
communication between qubits.
Finally, an effective method still lacks for refocusing the
dynamical-phase shifts of the qubits to realize the desired
quantum operations.

Here, we propose an effective scheme for coupling any pair of
superconducting qubits without direct interaction between them by
letting these be sequentially connected to a large CBJJ that acts
as a data bus. The qubit in Ref.~\cite{Strauch03} is a CBJJ, while
here we consider charge qubits. Here, a large CBJJ acts only as
the information bus between the qubits.
Also, in contrast to Ref.~\cite{Blais03}, in the present circuit
any chosen qubit can be coupled to and decouple from the bus by
switching on and off its Josephson energy. The bias current
applied to the bus is fixed during the operations, and the
dynamical-phase shifts of the qubits can be conveniently refocused
by properly setting the free-evolution times of the bus.
Therefore, an entanglement between distant qubits can be created
in a controllable way for realizing quantum computation, and also
for testing Bell's inequality.
Its experimental realizability is also briefly discussed.

{\it Model.---\/} Without loss of generality, we consider the
simplest network sketched in Fig. 1.
\begin{figure}
\vspace{1.8cm}
\includegraphics[width=12cm]{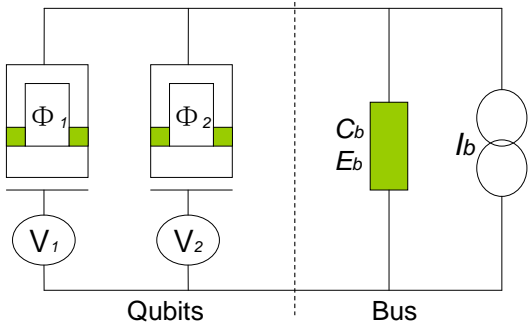}
\vspace{-5.5cm} \caption{A pair of SQUID-based charge qubits,
located on the left of the dashed line, coupled to a large CBJJ on
the right, which acts as an information bus. The circuit is
divided into two parts, the qubits and the bus. The dashed line
only indicates a separation between these. The controllable gate
voltage $V_k\,(k=1,2)$ and external flux $\Phi_k$ are used to
manipulate the qubits and their interactions with the bus. The bus
current remains fixed during the operations.}
\end{figure}
It can be easily modified to include arbitrary qubits. Each qubit
consists of a gate electrode of capacitance $C_g$ and a
single-Cooper-pair box with two ultrasmall identical Josephson
junctions of capacitance $c_J$ and Josephson energy
$\varepsilon_J$, forming a superconducting quantum interference
device (SQUID) ring threaded by a flux $\Phi$ and with a gate
voltage $V$. The superconducting phase difference across the $k$th
qubit is represented by $\phi_k$, $k=1,2$.
The large CBJJ has capacitance $C_{b}$, phase drop $\phi_{b}$,
Josephson energy $E_b$, and a bias current $I_b$.
The qubit is assumed to work in the charge regime with $k_BT\ll
E_J\ll E_C\ll\Lambda$, wherein quasi-particle tunnelling or
excitations are effectively suppressed. Here,
$k_B,\,T,\,\Lambda,\,E_C$, and $E_J$ are the Boltzmann constant,
temperature, superconducting gap, charging and Josephson coupling
energies of the qubit, respectively.
The present mechanism of quantum manipulation is significantly
different from Refs.~\cite{Makhlin99,Blais03,You02}, although the
circuits appear to be similar. The Hamiltonian for the circuit in
Fig. 1 is
\begin{eqnarray}
\hat{H}&=&\sum_{k=1}^2\left\{\frac{4e^2}{2C_k}\left[\hat{n}_k-n_g^{(k)}\right]^{2}-
E_J^{(k)}\cos\left[\hat{\phi}_{k}-\frac{C_g^{(k)}}{C_k}\hat{\phi}_{b}\right]\right\}
\nonumber\\
&+&\frac{\hat{Q}^2_b}{2\tilde{C}_b}-E_{b}\cos\hat{\phi}_b
-\frac{\Phi_0I_b}{2\pi}\,\hat{\phi}_b,
\end{eqnarray}
with $[\hat{\phi}_k,\hat{n}_k]=i$. Here,
$\hat{Q}_b=2\pi\hat{p}\,_b/\Phi_0$ is the operator of charges on
the CBJJ and $[\hat{\phi}_b,\hat{p}\,_b]=i\hbar$.
$C_J^{(k)}=2c_{J}^{(k)}$,\,
$E_J^{(k)}=2\varepsilon_J^{(k)}\cos(\pi\Phi_{k}/\Phi_0)$,\,$C_k=C_g^{(k)}+C_J^{(k)}$,\,
$n_{g}^{(k)}=C_g^{(k)}V_{k}/(2e)$,\, and $
\tilde{C}_b=C_b+\sum_{k=1}^2C_J^{(k)}C_g^{(k)}/C_k$.
$\Phi_0=h/(2e)$ and $n_k$ are the flux quantum and the excess
number of Cooper pairs in the superconducting box of the $k$th
qubit, respectively.
When the applied gate voltage $V_k$ is set near the degeneracy
points ($n_{g}^{(k)}=(l+1/2),\,l=0,1,2,...$), then only the two
lowest-energy
charge states, $|n_k=0\rangle=|\!\uparrow_k\rangle$ and $|n_k=1\rangle=|%
\!\downarrow_k\rangle$, play a role.
The large CBJJ works in the phase regime and describes the motion
of a ``particle" with mass $m=\tilde{C}_b(\Phi_0/2\pi)^2$
in a potential $U(\phi_b)=-E_{b}(\cos\phi_b+I_b\phi_b/I_c)$ with $%
I_c=2\pi E_{b}/\Phi_0$. For $I_b<I_c$, there exist a series of
minima of $U(\phi_b)$. Near these points, $U(\phi_b)$ approximates
a harmonic oscillator potential with characteristic frequency
$\omega_b=\sqrt{8E^{(b)}_CE_b/\hbar^2}\,\left[1-(I_b/I_c)^2\right]^{1/4},\,E^{(b)}_{C}=e^2/(2\tilde{C}_b)$.
The approximate number of metastable quantum bound
states~\cite{Clarke88} is $
N_s=2^{3/4}\sqrt{E_{b}/E^{(b)}_{C}}(1-I_b/I_c)^{5/4}$.
For a low bias current, the dynamics of the CBJJ can be safely
restricted~\cite{martinis02} to the Hilbert space spanned by the
two lowest states of this data bus: $|0_b\rangle$ and
$|1_b\rangle$.

The eigenenergy $\hbar\omega_{LC}$ of the bus used in
\cite{Makhlin99} is much higher than that of the qubits.
Therefore, adiabatically eliminating such a bus yields a direct
interbit coupling.
However, the energy scale of our proposed data bus (i.e., the CBJJ
oscillator) is $\omega_b/2\pi\,\sim 10$ GHz~\cite{martinis02},
which is of the same order of the Josephson energy (e.g., $%
E_J/h\,\sim 13$ GHz~\cite{Tsai99}). Therefore, the quanta in the
present bus can be excited or de-excited when the qubit is
operated.
The Hamiltonian (1) clearly shows that the coupling between the
chosen $k$th qubit and the bus can be turned on and
off~\cite{ref1}, when the threaded flux $\Phi_k$ differs from or
equals to $(l'+1/2)\Phi_0,\,l'=0,1,2,...$. For simplicity,
hereafter we let $l,\,l'=0$.
Two qubits can be indirectly coupled by independently interacting
with the bus sequentially when exciting/de-exciting the
vibrational quanta of the bus.
Under the usual rotating-wave approximation, the dynamics for such
a coupling mechanism can be described by the following effective
Hamiltonian
\begin{eqnarray}
&&\hat{H}_{kb}=\hat{H}_k+\hat{H}_b+i\lambda_k
\left[\hat{\sigma}^{(k)}_+\hat{a}-\hat{\sigma}^{(k)}_-\hat{a}^\dagger\right],\\
&&\hat{H}_k=\frac{E_J^{(k)}}{2}\hat{\sigma}^{(k)}_z- \frac{\delta
E_C^{(k)}}{2}\hat{\sigma}^{(k)}_x,\,\,\,
\hat{H}_b=\hbar\omega_b\left(\hat{n}+\frac{1}{2}\right),\nonumber
\end{eqnarray}
with $\hat{a}=\left[\sqrt{
\tilde{C}_b\omega_b/\hbar}\left(\frac{\Phi_0}{2\pi}\right)\hat{\phi}_b+
i\left(\frac{2\pi}{\Phi_0}\right)\hat{p}_b/\sqrt{\hbar\omega_b\tilde{C}_b}\,\right]/\sqrt{2}$
and $\hat{n}=\hat{a}^\dagger\hat{a}$ being the Boson operators of
the bus. Here, $\delta E_C^{(k)}=2e^2(1-2n^{(k)}_{g})/C_k$,
$\lambda_k=\zeta_k\cos(\pi\Phi_k/\Phi_0),\,
\zeta_k=\varepsilon^{(k)}_JC_g^{(k)}\pi\sqrt{2\hbar}/(C_k\Phi_0\sqrt{\tilde{C}_b\omega_b}\,)$.
The pseudospin operators: $\hat{\sigma}^{(k)}_z=|1_k\rangle\langle
1_k| -|0_k\rangle\langle 0_k|$,\,
$\hat{\sigma}^{(k)}_+=|1_k\rangle\langle 0_k|$, and
$\hat{\sigma}^{(k)}_-=|0_k\rangle\langle 1_k|$ are defined in the
subspace spanned by the logic states:\, $
|0_k\rangle=(|\!\!\downarrow_k\rangle+|\!\!\uparrow_k\rangle)/\sqrt{2}$
and
$|1_k\rangle=(|\!\!\downarrow_k\rangle-|\!\!\uparrow_k\rangle)/\sqrt{2}$.
Only the single-quantum transition process, approximated to
first-order in $\hat{\phi}_b$, is considered during the expansion
of the cosine-term of the Hamiltonian (1), as the fluctuation of
$\phi_b$ is very weak. In fact,
$C_g^{(k)}\sqrt{\langle\,\hat{\phi}^2_b\,\rangle}/C_k\,
\lesssim\,10^{-2}\ll\,1$, for typical experimental
parameters~\cite{Tsai99,Tsai03,martinis02}: $C_b\sim
6\,p$F,\,$\omega_b/2\pi\,\sim 10$ GHz, and
$C_g^{(k)}/C_J^{(k)}\sim 10^{-2}$.
Once the bias current $I_b$ is properly set up beforehand, various
dynamical evolutions can be induced by selecting the applied flux
$\Phi_k$ and the gate voltage $V_k$.
Considering two extreme
cases, the strongest coupling ($\Phi_k=0$) and the decoupling
($\Phi_k=\Phi_0/2$), several typical realizable evolutions deduced
from the Hamiltonian (2) are given in table I. There,
$\hbar\Delta_k=\varepsilon_k-\hbar\omega_b,\,\varepsilon_k=\sqrt{[2\varepsilon^{(k)}_J]^2+[\delta
E^{(k)}_C]^2}$.
\begin{table}
\begin{tabular}{c|c}
\hline\hline \cline{1-2}
Controllable Parameters&  Evolutions        \\
\hline
 $V_k=e/C_g^{(k)},\,\,\,\,\Phi_k=\Phi_0/2$
  &
 $\hat{U}_{0}(t)$  \\
 \hline
 $V_k\neq e/C_g^{(k)},\,\,\,\,\Phi_k=\Phi_0/2$&
  $\hat{U}^{(k)}_{1}(t)$  \\
\hline
$V_k=e/C_g^{(k)},\,\,\,\,\Phi_k=0,\,\,\,\,\hbar\omega_b=2\varepsilon^{(k)}_J$
&
$\hat{U}^{(k)}_{2}(t)$\\
\hline $V_k\neq e/C_g^{(k)},\,\,\,\,\Phi_k=0,\,\,\,\,
2\zeta_k\sqrt{n+1}\ll\hbar\Delta_k$ &
 $\hat{U}^{(k)}_{3}(t)$   \\
\hline
\end{tabular}
\caption{\label{tab1} Typical settings of the controllable
experimental parameters ($V_k$ and $\Phi_k$) and the corresponding
time evolutions $\hat{U}_j(t)$ of the qubit-bus system. Here,
$C^{(k)}_g$ and $2\varepsilon_J^{(k)}$ are the gate capacitance
and the maximal Josephson energy of the $k$th SQUID-based charge
qubit. $\zeta_k$ is the maximum strength of the coupling between
the $k$th qubit with energy $\varepsilon_k$ and the bus of
frequency $\omega_b$. The detuning between the qubit and the bus
energies is $\hbar\Delta_k=\varepsilon_k-\hbar\omega_b$. $n=0,1$
is occupation number for the number state $|n\rangle$ of the bus.
The various time-evolution operators are:
$\hat{U}_{0}(t)=\exp(-it\hat{H}_b/\hbar)$,\,$\hat{U}^{(k)}_{1}(t)=\exp[-it\delta
E_C^{(k)}\hat{\sigma}_x^{(k)}/(2\hbar)]\otimes
\hat{U}_{0}(t)$,\,$\hat{U}_2^{(k)}=\hat{A}(t)\cos\hat{\lambda}_n|0_k\rangle\langle
0_k|-(\sin\hat{\lambda}_n)\hat{a}/\sqrt{\hat{n}+1}|0_k\rangle\langle
1_k|+
\hat{a}^\dagger\sin\hat{\xi}_n/\sqrt{\hat{n}}|1_k\rangle\langle
0_k|+\cos\hat{\xi}_n|0_k\rangle\langle 0_k|$, and
$\hat{U}_3^{(k)}(t)=\hat{A}(t)\exp\{-it\zeta_k^2[|1_k\rangle\langle
1_k|(\hat{n}+1)-|0_k\rangle\langle
0_k|\hat{n}]/(\hbar\Delta_k)\}$, with
$\hat{A}(t)=\exp[-it(2\hat{H}_b+E_J^{(k)}\hat{\sigma}_z^{(k)})/(2\hbar)]$,
$\hat{\lambda}_n=2\zeta_kt\sqrt{\hat{n}+1}/\hbar$, and
$\hat{\xi}_n=2\zeta_kt\sqrt{\hat{n}}/\hbar$.}
\end{table}

{\it Quantum gates.---}
The physical characteristic (e.g., the eigenfrequency) of the bus
in the present circuit does not need to be changed, once it is set
up beforehand. It still undergoes a free evolution $\hat{U}_0(t)$
ruled by a non-zero $\hat{H}_b$ during the operational delay,
i.e., the time interval when the qubits do not evolve because
their Hamiltonians are temporarily set to zero (when
$\Phi_k=\Phi_0/2,\,V_k=e/C^{(k)}_g$). Before and after the desired
operations, the bus should remain in its ground state
$|0_b\rangle$. In principle, the time-evolutions listed in table I
are sufficient to implement any desired operation for manipulating
the quantum information stored in the present circuit. In fact,
any single-qubit rotation, including the typical Hadamard gate:
$\hat{H}^{(k)}_g=[\hat{\sigma}^{(k)}_z+\hat{\sigma}^{(k)}_++\hat{\sigma}^{(k)}_-]/\sqrt{2}$,
on the $k$th qubit can be easily realized by selectively using
$\hat{U}^{(k)}_{1}(t)$ and $\hat{U}^{(k)}_{3}(t)$. Any
single-qubit operation is not influenced by the free evolution of
the bus during the time delay between successive operations.

In order to realize two-qubit gates, we must be able to couple
distant qubits via their sequential interactions to the bus. We
set the bias current $I_b$ such that
$\hbar\omega_b=2\varepsilon^{(1)}_{J}$ and then perform a
three-step process.
First, we couple the first (control) qubit to the bus by switching
off its applied flux $\Phi_1$ and produce the evolution
$\hat{U}^{(1)}_{2}(t_1)$. After a duration $t_1$ determined by
$\sin(2\zeta_1t_1/\hbar)=-1$, the control qubit is decoupled from
the bus. This process implements the evolutions:
$|0_b\rangle|0_1\rangle\rightarrow|0_b\rangle|0_1\rangle$ and
$|0_b\rangle|1_1\rangle\rightarrow
e^{-i\omega_bt_1}|1_b\rangle|0_1\rangle$.
Next, we let the second (target) qubit work at a non-degenerate
point ($V_2\neq e/C^{(2)}_{g}$) and couple it to the bus by
switching off its applied flux $\Phi_2$.
After a duration $t_2$ determined by
\begin{eqnarray}
\sin\left(\frac{\zeta^2_2t_2}{\hbar^2\Delta_2}\right)=
\cos\left(\frac{\varepsilon_2t_2}{2\hbar}+\frac{\zeta^2_2t_2}{2\hbar^2\Delta_2}\right)=1,
\end{eqnarray}
the target qubit is backed to its degenerate point
($V_2=e/C^{(2)}_g$) and decoupled from the bus. This sequence of
operations generate the evolutions:
\begin{eqnarray*}
\left\{
\begin{array}{ll}
|0_b\rangle|0_2\rangle\rightarrow
e^{-i\xi}|0_b\rangle|0_2\rangle,\,\,
|0_b\rangle|1_2\rangle\rightarrow
e^{-i\xi}|0_b\rangle|1_2\rangle,\\
|1_b\rangle|0_2\rangle\rightarrow
i\,e^{-i(\xi+\omega_bt_g)}\left(\cos\eta_2|1_b\rangle|0_2\rangle+\sin\eta_2|1_b\rangle|1_2\rangle\right),\\%
|1_b\rangle|1_2\rangle\rightarrow
i\,e^{-i(\xi+\omega_bt_g)}\left(\sin\eta_2|1_b\rangle|0_2\rangle-\cos\eta_2|1_b\rangle|1_2\rangle\right),
\end{array}
\right.
\end{eqnarray*}
with
$\xi=\omega_b(\tau_1+t_2+\tau_2)/2+\zeta^2_it_2/(2\hbar^2\Delta_2)$,
and $t_g=\sum_{s=1}^3t_s+\sum_{s=1}^2\tau_s$.
Finally, we couple again the control qubit to the bus and perform
the evolution $\hat{U}^{(1)}_{2}(t_3)$ with
$\sin(2\zeta_1t_3/\hbar)=1$, yielding evolutions:
$|0_b\rangle|0_1\rangle\rightarrow|0_b\rangle|0_1\rangle$ and
$|1_b\rangle|0_1\rangle\rightarrow
e^{-i\omega_bt_3}|0_b\rangle|1_1\rangle$.
In practice, the free evolutions $\hat{U}_{0}(\tau_1)$ and
$\hat{U}_{0}(\tau_2)$ exist during the time delays between the
first (second) and second (third) pulses. If the delays are
further set accurately such that the total duration $t_g$
satisfies the condition $\sin\omega_bt_g=1$, then the above
three-step process with two delays yields a quantum operation
$
\hat{U}(t_g)=\hat{U}^{(1)}_{2}(t_3)\hat{U}_{0}(\tau_2)\hat{U}^{(2)}_{3}(t_2)
\hat{U}_{0}(\tau_1)\hat{U}^{(1)}_{2}(t_1)=\exp(-i\xi)|0_b\rangle\langle
0_b|\otimes\hat{U}^{(12)}_d(\beta_2),
$
with
\begin{eqnarray}
\hat{U}^{(12)}_d(\beta_2)=\left(
\begin{array}{cccc}
1&0&0&0\\
0&1&0&0\\
0&0&\cos\beta_2&\sin\beta_2\\
0&0&\sin\beta_2&-\cos\beta_2
\end{array}
\right)
\end{eqnarray}
being a universal two-qubit gate. Here,
$\cos\beta_2=2\varepsilon_J^{(2)}/\varepsilon_2$. This gate can
produce entanglement between qubits and also realize any quantum
computation, accompanied by single-qubit rotations.

{\it Testing Bell's inequality.---} Entanglement is a key
ingredient for computational speedup in quantum computation.
Historically, Bell's inequalities were seen as an entanglement
test: its violation implies that entanglement must exist.
For a two-qubit entangled state $|\psi_e\rangle$, the Clauser,
Horne, Shimony and Holt (CHSH) form  of Bell's inequality : $
f(|\psi_e\rangle)\leq 2$\, is usually tested by experimentally
measuring the CHSH function
$f(|\psi_e\rangle)=|E(\theta_1,\theta_2)+E(\theta_1^{\prime},%
\theta_2)+E(\theta_1,\theta_2^{\prime})
-E(\theta_1^{\prime},\theta_2^{\prime})|.
$
Here, $\theta_k$ are controllable classical variables and
$E(\theta_1,\theta_2)$ is the correlation for the outcomes of
separately projected measurements of two qubits. A number of
experimental tests~\cite{Rowe01} of Bell's inequality have already
been performed by using entangled photons and atoms. We now show
that a desired entangled state can be created in a repeatable way
and thus Bell's inequality can also be tested experimentally by
using this circuit.

We begin with an initial state
$|\psi_0\rangle=|0_b\rangle|\!\!\downarrow_1\rangle|\!\!\downarrow_2\!\rangle$=$|0_b\rangle
\otimes\left(|0_1\rangle+|1_1\rangle\right)\otimes
\left(|0_2+|1_2\rangle\right)/2$ with two qubits decoupled from
the bus but working at their non-degenerate points (i.e., $V_k\neq
e/C_g^{(k)}$). After applying a Hadamard gate $\hat{H}^{(2)}_g$ to
the second qubit, the system evolves to the state
$|\psi_1\rangle=|0_b\rangle
\otimes\left(|0_1\rangle+|1_1\rangle\right)\otimes
|1_2\rangle/\sqrt{2}$. %
The desired two-qubit entangled state is then generated as
\begin{eqnarray}
|\psi^{(12)}_d(\theta_1,\theta_2,\beta_2)\rangle
=\hat{U}^{(1)}_{1}(\theta_1)\hat{U}^{(2)}_{1}(\theta_2)\hat{U}_d^{(12)}(\beta_2)
|\psi_1\rangle,
\end{eqnarray}
with
$\hat{U}^{(k)}_{1}(\theta_k)=\exp\left[i\theta_k\hat{\sigma}^{(k)}_x/2\right],\,\theta_k=\delta
E_C^{(k)}t_k/\hbar$.
The corresponding correlation function is
$E(\theta_1,\theta_2,\beta_2)=\langle\psi^{(12)}_d(\theta_1,\theta_2,\beta_2)
|\hat{\sigma}^{(1)}_z\otimes\hat{\sigma}^{(2)}_z
|\psi^{(12)}_d(\theta_1,\theta_2,\beta_2)\rangle
=\sin\beta_2(\sin\theta_1\sin\theta_2$$-$$\sin\beta_2\cos\theta_1\cos\theta_2)$.
For certain chosen sets of angles: $\theta_k=\{-\pi/8,\,3\pi/8\}$,
the CHSH function becomes
\begin{eqnarray}
f(|\psi^{(12)}_d(\beta_2)\rangle)
=\sqrt{2}\,|\sin\beta_2(\sin\beta_2+1)|.
\end{eqnarray}
It is easy to numerically check that Bell's inequality,
$f(|\psi^{(12)}_d(\beta_2)\rangle)\leq 2$, is violated when
$E^{(2)}_{J}/\delta E^{(2)}_{C}\,<\,0.776$, which can be easily
satisfied for this charge-qubit system.

Experimentally, the above procedure can be repeated many times at
each of the four sets of angles and thus the correlation function
$
E(\theta_1,\theta_2,\beta_2)=[N_{\rm same}(\theta_1,\theta_2)-N_{\rm diff}(\theta_1,%
\theta_2)]/N_{\rm tot}$, with $N_{\rm same}(\theta_1,\theta_2)$
($N_{\rm diff}(\theta_1,\theta_2)$) being the number of events
with two qubits being found in the same (different) logic states
and
$N_{\rm tot}=N_{\rm same}(\theta_1,\theta_2)+N_{\rm diff}(%
\theta_1,\theta_2)$ being the total experimental times for the
same $\theta_1$ and $\theta_2$.
Finally, Bell's inequality can be tested by calculating the
experimental CHSH function:
$f(|\psi^{(12)}_d(\beta_2)\rangle)=|E(\theta_1,\theta_2,\beta_2)+E(\theta_1^{\prime},%
\theta_2,\beta_2)+E(\theta_1,\theta_2^{\prime},\beta_2)
-E(\theta_1^{\prime},\theta_2^{\prime},\beta_2)|$.

{\it Discussion.---} Two types of noise, fluctuations of the
applied gate voltage $V_k$ and bias current $I_b$, must be
considered in the present qubit-bus system.
For simplicity, these two environmental noises are treated as two
separate Boson baths with Ohmic spectral densities and assumed to
be weakly coupled to the qubit and CBJJ, respectively.
The Hamiltonian of the $k$th qubit coupling to the bus, containing
these fluctuations, can be written as
\begin{eqnarray}
\hat{H}&=&\hat{H}_{kb}+\sum_{j=1,2}\sum_{\omega_j}\hbar \omega
_j\hat{a}_{\omega_j}^{\dagger}\hat{a}_{\omega_j}
-\frac{eC^{(k)}_g}{C_k}\hat{\sigma}^{(k)}_z(\hat{R}_{1}+\hat{R}^\dagger_1)\nonumber\\
&-&\sqrt{\frac{\hbar}{2\widetilde{C}_{b}\omega
_b}}\left(\hat{a}^{\dagger}\hat{R}_{2}+\hat{a}\hat{R}_{2}^{\dagger}\right),\,\hat{R}_{j}=\sum_{\omega_{j}}g_{\omega
_{j}}\hat{a}_{\omega_{j}}.
\end{eqnarray}%
Here, $\hat{a}_{\omega_j}$ is the Boson operator of the $j$th bath
and $g_{\omega_j}$ its coupling strength.
The relaxation and decoherence rates of our qubit-bus system can
also be calculated by using the well established Bloch-Redfield
formalism~\cite{Storcz03}.
Under the usual secular approximation, the relaxation and
decoherence rates are characterized~\cite{ref2} by two
dimensionless coupling parameters,
$\alpha_V=(C^{(k)}_g/C_k)^2R_V/R_K$ and $\alpha_I={\rm
Re}(Y_I)/(\tilde{C}_b\omega_b)$, which describe the couplings of
the voltage fluctuations to the qubit and the bias-current
fluctuations to the bus, respectively. Here, $R_K=h/e^2\approx
25.8$~k$\Omega$ is the quantum of resistance, $R_V$ is the Ohmic
resistor of the voltage, and ${\rm Re}(Y_I)$ is the dissipative
part of the admittance of the current bias.
If the qubit decouples from the bus, $\alpha_V$ ($\alpha_I$)
characterizes the decoherence and relaxation of the qubit (bus).
It has been estimated in Ref.~\cite{Makhlin01} that the
dissipation for a single SQUID-based charge qubit is sufficiently
weak ($\alpha_V\approx 10^{-6}$), which allows, in principle, for
$10^{6}$ coherent single-qubit manipulations. However, for a
single CBJJ the dimensionless parameter $\alpha_I$ only reaches
$10^{-3}$ for typical experimental parameters~\cite{martinis02}:
$1/{\rm Re}(Y_I)\sim 100~\Omega$, $C_b\sim~6~p$F,
$\omega_b/2\pi\,\sim 10~$GHz.
This implies that the quantum coherence of the present qubit-bus
system is mainly limited by the bias-current fluctuations.
Fortunately, the impedance of the above CBJJ can be engineered to
be $1/{\rm Re}(Y_I)\sim 560~ $k$\Omega$ \cite{martinis02}. This
lets $\alpha_I$ reach up to $10^{-5}$ and allow about $10^{5}$
coherent manipulations of the qubit-bus system.

In summary, we have proposed a scheme for coupling two SQUID-based
charge qubits by sequentially using their interactions with a
common large Josephson junction biased by a fixed current. Each
interaction is tunable by controlling the external flux applied to
the chosen SQUID-based charge qubit.
The proposed circuit allows the possibility of implementing
elementary quantum logic operations, including arbitrary
single-qubit gates and universal two-qubit gates.
The created two-qubit entangled states can be used to test Bell's
inequality.

\begin{acknowledgments}
We acknowledge useful discussions with Drs. J.Q. You, J.S. Tsai,
and X. Hu, and the support of the US NSA and ARDA under AFOSR
contract No. F49620-02-1-0334, and the NSF grant No. EIA-0130383.
\end{acknowledgments}


\vspace{-0.7cm}

\begin{thebibliography}{99}
\vspace{-0.8cm}
%
\bibitem{Makhlin01} Y. Makhlin, G. Sch\"on and A. Shnirman, Rev. Mod. Phys. {\bf 73}, 357
(2001).
%
\bibitem{Tsai99} Y. Nakamura, Yu.A. Pashkin, and J.S. Tsai, Nature
{\bf 398}, 786 (1999); K.W. Lehner {\it et al}., Phys. Rev. Lett.
{\bf 90}, 027002 (2003).
%
\bibitem{Mooij99} J.E. Mooij {\it et al.}, Science {\bf 285},
1036 (1999); {\bf 290}, 773 (2000); E. Il'ichev {\it et. al.},
Phys. Rev. Lett. {\bf 91}, 097906 (2003).
%
\bibitem{Vion02} D. Vion {\it et al.}, Science {\bf 296}, 886 (2002).
%
\bibitem{Yu02} Y. Yu {\it et al.}, Science {\bf 296}, 889 (2002).
%
\bibitem{martinis02} J.M. Martinis {\it et al.}, Phys. Rev. Lett. {\bf 89}, 117901 (2002).
%
%
\bibitem{Makhlin99} Y. Makhlin, G. Sch\"on and A. Shnirman, Nature {\bf 398}, 305 (1999).
%
\bibitem{Tsai03}Yu.A. Pashkin {\it et al.}, Nature
\textbf{421}, 823 (2003); T. Yamamoto {\it et al.}, Nature
\textbf{425}, 941 (2003).
%
\bibitem{Averin03} D.V. Averin and C.
Bruder, Phys. Rev. Lett. {\bf 91}, 057003 (2003).
%
\bibitem{You02}
J.Q. You, J.S. Tsai, and F. Nori, Phys. Rev. Lett. {\bf 89},
197902 (2002); Phys. Rev. B {\bf 68}, 024510 (2003).
%
%
\bibitem{Blais03} A. Blais, A.M. van den Brink, and A.M. Zagoskin,
Phys. Rev. Lett. {\bf 90}, 127901 (2003).
%
\bibitem{Strauch03}
A.J. Berkley {\it et al.}, Science {\bf 300}, 1548 (2003); F.W.
Strauch {\it et al.}, Phys. Rev. Lett. {\bf 91}, 167005 (2003).
%
%
%
\bibitem{Falci03}
F. Plastina and G. Falci, Phys. Rev. B {\bf 67}, 224514 (2003).

\bibitem{Zhu03}
S.L. Zhu, Z.D. Wang, and K. Yang, Phys. Rev. A {\bf 68}, 034303
(2003); C.-P. Yang and S.-I Chu, {\it ibid.} {\bf 67}, 042311
(2003); J.Q. You and F. Nori, Phys. Rev. B {\bf 68}, 064509
(2003).

\bibitem{Clarke88}
J. Clarke {\it et al.}, Science {\bf 239}, 992 (1988);
%
J.M. Martinis, M.H. Devoret, and J. Clarke, Phys. Rev. B {\bf 35},
4682(1987).
%
\bibitem{Storcz03}
M. Storcz and F.K. Wilhelm, Phys. Rev. A {\bf 67}, 042319 (2003);
E. Paladino {\it et al.}, Phys. Rev. Lett. {\bf 88}, 228304
(2002); U. Weiss, {\it Quantum Dissipative systems}, 2nd ed.
(World Scientific, Singapore, 1999).
%
\bibitem{CZ95} J.I. Cirac and P. Zoller, Phys. Rev. Lett. {\bf 74}, 4091
(1995); L.F. Wei, S.Y. Liu, and X.L. Lei, Phys. Rev. A {\bf 65},
062316 (2002); Opt. Comm. {\bf 208}, 131 (2002).
%
\bibitem{ref1}
Indeed, such a possibility requires that the Josephson energies of
two junctions in a SQUID-loop to be equal or have a very small
relative difference $|\delta \varepsilon_J/\varepsilon_J|$ in
their coupling energies. Experimentally~\cite{Rouse95}, this has
been reached with $|\delta \varepsilon_J/\varepsilon_J|\sim 1\%$.
This implies that another interaction term $\delta H\approx \delta
\varepsilon_J\sin(\pi\Phi_k/\Phi_0)\sin\left[\hat{\phi}_k-C_g^{(k)}\hat{\phi}_b/C_k\right]$,
due to the different critical currents of the two junctions, can
be safely neglected and thus the system can be described by the
Hamiltonian (1).
\bibitem{Rouse95}
See, e.g., R. Rouse, S. Han, and J.E. Lukens, Phys. Rev. Lett.
{\bf 75}, 1614 (1995).

\bibitem{Rowe01}
M.A. Rowe {\it et al.}, Nature {\bf 409}, 791 (2001); G. Weihs
{\it et al.}, Phys. Rev. Lett. {\bf 81}, 5039 (1998).
%
%
\bibitem{ref2} Only a few lower-energy
eigenstates of $\hat{H}_{kb}$, i.e., the ground state
$|g\rangle$$=$$|-_k,0_b\rangle$ and the first doublet,
$|u\rangle$$=$$\cos\chi_k|+_k,0_b\rangle$$-$$i\sin\chi_k|-_k,1_b\rangle$
and
$|v\rangle$$=$$-i\sin\chi_k|+_k,0_b\rangle$$+$$\cos\chi_k|-_k,1_b\rangle$,
are involved in our calculations. Their corresponding eigenvalues
are $\hbar\omega_g$, $\hbar\omega_{u}$, and $\hbar\omega_{v}$,
respectively. Here, $|\pm_k\rangle$ ($|0_b\rangle,|1_b\rangle$)
are the eigenstates of $\hat{H}_k$ ($\hat{H}_b$), and
$\cos\chi_k$$=$$\sqrt{(\rho_k-\delta_k)/(2\rho_k)},\,
\rho_k$$=$$\sqrt{\delta_k^2+4\lambda^2_k},\,\delta_k$$=$$\varepsilon_k$$-$$\hbar\omega_b$.
This simplifies the calculation of the rates of decoherence and
relaxation. For example, the decoherence rate of the superposition
of states $|u\rangle$ and $|v\rangle$ can be estimated as
$\gamma_{uv}\sim\alpha_V\,A_V+\alpha_I\,A_I$, with
$A_V$$=$$B_1\sin^2\alpha_k+B_2\cos^2\alpha_k,\,
A_I$$=$$\Omega_{ug}\sin^2\chi_k+\Omega_{vg}\cos^2\chi_k$,
$B_1=4\cos^2(2\chi_k)(2k_BT)/\hbar$$+$$2\Omega_{vu}\sin^2(2\chi_k)$,
and $B_2=\Omega_{ug}\cos^2\chi_k+\Omega_{vg}\sin^2\chi_k$.
%
Also, $\Omega_{ug}=\omega_{ug}\coth[\hbar\omega_{ug}/(2k_bT)]$,
$\Omega_{vg}=\omega_{vg}\coth[\hbar\omega_{vg}/(2k_bT)]$,
$\Omega_{vu}$$=$$\omega_{vu}\coth[\hbar\omega_{vu}/(2k_bT)]$ and
$\omega_{vu}$$=$$\omega_{v}$$-$$\omega_{u}$, etc.
%
Specifically, for the decoupling case with $\sin\chi_k=0$, the
qubit and bus independently dephase with the rate
$\gamma_{+-}\sim\alpha_V\{4\sin^2\alpha_k2k_BT/\hbar+
\omega_{+-}\cos^2\alpha_k\coth[\hbar\omega_{+-}/(2k_bT)]\}$ and
$\gamma_{10}\sim\alpha_I\omega_b\coth[\hbar\omega_b/(2k_BT)]$.
%

\end{thebibliography}
\end{document}